\documentclass[twocolumn,english]{article}
\usepackage[T1]{fontenc}
\usepackage[latin9]{inputenc}
\usepackage{url}
\usepackage{amsmath}
\usepackage{amssymb}

\makeatletter

\providecommand{\tabularnewline}{\\}

\newenvironment{lyxlist}[1]
{\begin{list}{}
{\settowidth{\labelwidth}{#1}
 \setlength{\leftmargin}{\labelwidth}
 \addtolength{\leftmargin}{\labelsep}
 }}
{\end{list}}

\makeatother

\usepackage{babel}
\begin{document}

\title{Processing XML for Domain Specific Languages}

\author{Tony Clark}
\maketitle
\begin{abstract}
XML is a standard and universal language for representing information.
XML processing is supported by two key frameworks: DOM and SAX. SAX
is efficient, but leaves the developer to encode much of the processing.
This paper introduces a language for expressing XML-based languages
via grammars that can be used to process XML documents and synthesize
arbitrary values. The language is declarative and shields the developer
from SAX implementation details. The language is specified and an
efficient implementation is defined as an abstract machine.
\end{abstract}

\section{Introduction}

XML is a standard and universal language for representing information.
It is used to represent information including: financial trading;
controlling robotic telescopes; clinical data; and, music. An XML
document consists of a tree of elements. Each element contains a tag,
some attribute name-value pairs and a sequence of child elements.
Leaf nodes may be unformatted text. 

In order for an XML document to be processed, it must conform to a
predefined format. The format defines a collection of tags that can
be used in the document, the attributes for an element with a given
tag and the rules of parent-child element composition. Such a format
defines a language and any XML document that conforms to the format
is written in the language. If the format is defined to support information
for a specific application domain (such as share prices or system
configuration) then it constitutes a \emph{domain specific language}
(DSL).

How should an XML document be processed? An application that processes
XML will need to read the document and translate it into some form
of useful information. This is often achieved using two approaches:
translate the XML into data that is then processed by the application;
translate the XML into calls on an application specific API. The first
approach can be thought of as a mapping from one data format to another
and the second as \emph{executing} the XML document. Sometimes a mixture
of the two is used.

In either case, working with XML involves reading a document and processing
the information in some way. There are two standard ways of processing
XML data:
\begin{lyxlist}{00.00.0000}
\item [{DOM}] A DOM processor \cite{DOM} translates the XML document into
a faithful in-memory tree and passes this data structure to the application.
The application can then traverse the tree and perform any appropriate
actions.
\item [{SAX}] A SAX framework \cite{SAX} traverses the XML document in
a predefined order and generates events for each type of tree-node
that it encounters. The application supplies the SAX framework with
an adapter that implements handlers for each event type. The handlers
perform application specific processing.
\end{lyxlist}
Both DOM and SAX processing will achieve the desired result. However,
there are significant drawbacks to the DOM approach since it requires
the complete XML tree to be represented in memory before application-specific
processing can take place. Firstly the XML document may be very large
so its representation in-memory may incur an unreasonable overhead.
Secondly, the DOM approach is not compatible with applications whose
life-cycle may be indefinite, for example interactive applications.

The SAX approach does not suffer from these drawbacks since the processing
of the XML data is interleaved with application specific event handlers.
Unfortunately, compared to DOM-based processing, writing a SAX processor
is complex since the SAX framework effectively flattens the XML tree
and generates a sequence of events.

SAX-based processing of a DSL involves recognizing sequences of events
that arise from a flattened XML document and performing actions that
either synthesize a data structure or make calls an an application
API. This processing is the same as the actions of a parser which
takes a description of a language (a grammar) and processes some input.
Given a suitable representation for XML grammars and an efficient
parsing engine then SAX processing of XML DSLs can be made both convenient
and efficient. 

This paper describes an approach to parsing XML grammars using a SAX
framework and shows how a standard LL(1) parsing technique can be
used to process XML documents. The grammar language is novel in that
it uses a convenient syntax in terms of parametric parsing rules and
can easily be implemented using an efficient parsing machine. The
language has been implemented and is available as part of the open-source
XMF system.

The paper is structured as follows: section \ref{sec:XML-Grammars}
describes a language for representing XML grammars; section \ref{sec:Specification}
specifies how the XML grammars process XML documents and synthesize
results; section \ref{sec:Implementation} defines a parsing machine
that is driven by an XML grammar and processes an XML document as
described in the specification; finally, section \ref{sec:Analysis}
reviews the paper and compares the results with similar systems.

\section{XML Grammars\label{sec:XML-Grammars}}

An XML grammar is a collection of rules. The rules specify a set of
legal XML documents; if document d is in the set of legal documents
for grammar g then g is satisfied by d. A grammar also specifies a
value for each XML document. If a document d satisfies grammar g with
value v then \emph{parsing} d with respect to g produces, or \emph{synthesizes},
value v.

The XMF system implements a parser for XML grammars. The grammars
are specified in a concrete language described in section \ref{sub:Example}.
The XMF-based grammar language is useful for humans, but long-winded
when describing precisely how the parsing mechanism works. Therefore,
section \ref{sub:Abstract-Syntax} defines an equivalent abstract
syntax for the grammar language that is used in the rest of the paper.

\subsection{Example\label{sub:Example}}

XMF implements XML grammars using a language that is based on BNF.
A grammar consists of rules that define non-terminals. The body of
a rule is a pattern that consists of element specifications (terminals),
rule calls (non-terminals), bindings and actions. The following is
an example of an XML grammar that processes a simple model language.
The models consist of packages, classes and associations. The rest
of this section describes the grammar in more detail.

\begin{small}
\begin{verbatim}
(1) @Grammar Models
(2)   Attribute ::=
(3)     <Attribute name type/>
(4)     { Attribute(name,type) }. 
      Class ::= 
        <Class name isAbstract id>
(5)       elements = ClassElement*
(6)     </Class> {
(7)       elements->iterate(e c = Class(name,isAbstract) | 
           c.add(e)) }.
(8)   ClassElement ::= Attribute | Operation.
      Operation ::= 
        <Operation name>
          as = Arg*
        </Operation> { Operation(name,as) }.
      Package ::=
        <Package name>
          elements = PackageElement*
        </Package> { 
          elements->iterate(e p = Package(name) | 
            p.add(e)) }.
      PackageElement ::= Package | Class | Assoc.
(9)   Assoc ::=
        <Association name>
          <End n1=name t1=type/>
          <End n2=name t2=type/>
        </Association> {
(10)      Association(n,End(n1,t1),End(n2,t2)) }.
    end
\end{verbatim}
\end{small}

\noindent The grammar is defined using the XML grammar DSL defined
by XMF and starts at line (1). Lines (2-4) are a typical example of
a grammar rule. The name of the rule is Attribute. The body specifies
that Attribute expects an XML element representing an attribute with
a name and a type. The variables name and type are bound to the values
of the corresponding XML attributes. Line (4) defines an action that
occurs after the XML element has been consumed. The action constructs
an instance of the XMF class Attribute and supplies the values of
name and type. Each component of a rule-body returns a value. The
value of the last component is that returned by a call of the rule.
In this case the rule returns a new attribute instance.

Line (5) is interesting because it shows a call of the rule ClassElement
and the use of the {*} decoration to specify that ClassElement should
be called repeatedly until it fails to be satisfied by the XML input.
The result of a component decorated with a {*} is a sequence of elements.

Line (8) is interesting because it shows how alternatives are specified
in a rule. A ClassElement is either an Attribute or an Operation.

Parsing starts with an initial rule and a tree (the root of the document).
Each rule element is processed in turn. Tree elements are consumed
each time an element specification (e.g. line 3) is encountered in
a rule. If the tags of the root element in the tree and the element
specification match then the root is consumed and the parse proceeds
with the child elements. If the comparison ever fails, and no further
choices are available, then the parse fails and no values are produced.

\subsection{Abstract Syntax\label{sub:Abstract-Syntax}}

In the rest of this paper we specify a parser for the XML grammar
language and give its implementation. The concrete language described
in the previous section is not really suitable for precise descriptions
of the specification and parsing machinery. Therefore, this section
gives an equivalent abstract syntax description of the essential features.

An abstract syntax for the grammar language is used as defined below
where $N$ is a set of names, E is a set of expressions, $\left\{ .\right\} $
is the power-set constructor, $\left[.\right]$ constructs a set of
sequences from an underlying type, $V$ is a set of values that can
be synthesized by a grammar and $t(P,...)$ denotes the set of all
terms with functor t constructed from the supplied sets P etc.\\

$\begin{array}{lcll}
g\in G & = & \left\{ C\right\}  & grammars\\
c\in C & = & N\times[N]\times B & clauses\\
b\in B & = &  & clause\; bodies\\
 &  & or(B,B) & disjunction\\
 & | & and(B,B) & conjunction\\
 & | & bind(\left[N\right],B) & binding\\
 & | & star(B) & repetition\\
 & | & empty & no\; elements\\
 & | & any & any\; element\\
 & | & ok & skip\\
 & | & text & raw\; text\\
 & | & call(N,[E]) & nonterminal\\
 & | & actions([E]) & synthesis\\
 & | & N\times\left\{ N\times N\right\} \times\Gamma\times B & element\: spec\\
\gamma\in\Gamma & = & E\mapsto B & guarded\; bodies\\
\rho\in\Phi & = & N\rightarrow V & environments\\
x\in X & = &  & XML\\
 &  & N\times\Phi\times[X] & element\\
 & | & text(S) & text
\end{array}$\\

\noindent A clause will be written $c(\tilde{v})\vartriangleright b$
where c is the name, v are the arguments and b is the body. A disjunction
will be written $b\mid b'$ and a conjunction $bb'$. A call will
be written $n(\tilde{e})$ and actions $\left[\tilde{e}\right]$.
Repetition will be written $b^{*}$. Bindings will be written $\tilde{n}=b$. 

An element specification is $(t,N,\gamma,b)$ which is to be interpreted
as follows: $t$ is a tag, $N$ is a set of names (actually a set
of name pairs to allow variables and attribute names to be different,
however we simplify this in definitions by assuming that they are
always the same) that specify the attributes to be bound when matching
against an XML element. The guarded bodies $\gamma$ is a function,
viewed as a set of pairs, associating boolean expressions with clause
body elements. The element $b$ is the else-clause.

Environments $\rho$ are just functions from names to values. They
will be extended in the normal way $\rho[n\mapsto v]$ and $\rho\oplus\rho'$
with shadowing on the right. The environment $\rho\backslash N$ is
the same as $\rho$ except that the domain is restricted to the set
of names N. 

Sequences of elements are written $\tilde{s}$ and are constructed
from the empty sequence {[}{]}, concatenation of sequences $\tilde{p}+\tilde{q}$
and consing $x:\tilde{s}$.

Expressions are used to represent guards in element specifications,
arguments in calls and synthesizing actions. An expression e may contain
variable references and denotes a value e($\rho).$ Sequences of actions
$\tilde{e}$ generalize naturally.

The Attribute rule body from the example concrete grammar described
in section \ref{sub:Example} is represented as follows using abstract
syntax (and a suitable action $e_{1}$):

\[
(Attribute,\left\{ name,type\right\} ,true,\left[e_{1}\right])
\]

\noindent The Operation rule body is:

\[
(Operation,\left\{ name\right\} ,true,[as]=Arg()^{*}\left[e_{2}\right])
\]

\subsection{Well Formedness Rules}

\begin{figure*}
\hfill{}
\[
\begin{array}{ccccc}
\dfrac{\begin{array}{c}
N\vdash b_{1}\\
N\vdash b_{2}
\end{array}}{N\cup(bound(b_{1})\cap bound(b_{2})\vdash b_{1}\mid b_{2}} & (Wor) & \qquad\qquad & \dfrac{\begin{array}{c}
N\vdash b_{1}\\
N\cup bound(b_{1})\vdash b_{2}
\end{array}}{N\cup bound(b_{1})\cup bound(b_{2})\vdash b_{1}b_{2}} & (Wand)\\
\\
\dfrac{\begin{array}{c}
free(\gamma)\subseteq N\cup N'\\
N\cup N'\vdash b\qquad\forall b\in ran(\gamma)\\
N\cup N'\vdash b
\end{array}}{N\vdash(n,N',\gamma,b)} & (Wel) &  & \dfrac{N\vdash b}{N\cup\tilde{n}\vdash bind(\tilde{n},b)} & (Wbind)\\
\\
N\vdash empty & (Wempty) &  & N\vdash any & (Wany)\\
\\
N\vdash text & (Wtext) &  & \dfrac{free(n(\tilde{e}))\subseteq N}{N\vdash n(\tilde{e})} & (Wcall)\\
\\
\dfrac{free(\left\{ \tilde{e}\right\} )\subseteq N}{N\vdash\left\{ \tilde{e}\right\} } & (Wsynth) &  & N\vdash ok & (Wok)
\end{array}
\]
\hfill{}

\protect\caption{Well-Formedness}
\label{fig:Well-formedness}
\end{figure*}

Not all syntactically correct grammar rules are meaningful. In order
for a rule to be correct it must conform to variable binding well-formedness
rules that require a variable to be bound before it can be referenced.
For example the following rule is not meaningful because the use of
disjunction means that the variable x cannot be guaranteed to be bound
in all cases:

\[
W()\vartriangleright(\left[x\right]=X()\mid\left[y\right]=Y())\: Z(x)
\]

\noindent The well-formedness rules depend on two functions that are
defined on the abstract syntax. The function $free:B\rightarrow\left\{ N\right\} $
is maps a rule element to a set of names that are freely referenced
in that element. The function $bound:B\rightarrow\left\{ N\right\} $
maps a rule element to the variable names that are bound by the element
and subsequently available once the element has been successfully
parsed.

A rule element $b$ is well formed when, given a context of bound
names $N$, the relationship $N\vdash b$ holds as defined in figure
\ref{fig:Well-formedness}. A rule $n(\tilde{n})\vartriangleright b$
is well formed when $\left\{ \tilde{n}\right\} \vdash b$ and a grammar
is well-formed when all of its rules are well-formed. 

Rule \emph{Wor} defines that names available outside a disjunction
must be bound by both parts of the disjunction. \emph{Wand} defines
that binding is sequential and cumulative. \emph{Wel} defines that
the names used in element specification guards must be in scope and
that the attributes are scoped over the guards and the child elements.
\emph{Wbind} defines that a binding element introduces names that
can be used in clause body element that occur subsequently. Both \emph{Wcall}
and \emph{Wsynth} require that freely referenced names must be bound.

\section{Specification\label{sec:Specification}}

\begin{figure*}
\hfill{}
\[
\begin{array}{ccccc}
\dfrac{\begin{array}{c}
g,b\vdash\tilde{x},\rho,\tilde{x}',v\end{array}}{g,b|b'\vdash\tilde{x},\rho,\tilde{x}',v} & (Sor_{1}) & \qquad\qquad & \dfrac{\begin{array}{c}
g,b\vdash\tilde{x},\rho,\tilde{x}',v\end{array}}{g,b'|b\vdash\tilde{x},\rho,\tilde{x}',v} & (Sor_{2})\\
\\
\dfrac{\begin{array}{c}
g,b_{1}\vdash\tilde{x},\rho_{1},\tilde{x}',v_{1}\\
g,b_{2}\vdash\tilde{x}',\rho_{2},\tilde{x}'',v_{2}
\end{array}}{g,b_{1}b_{2}\vdash\tilde{x},\rho_{1}\oplus\rho_{2},\tilde{x}'',v_{2}} & (Sand) &  & \dfrac{g,b\vdash\tilde{x},\rho,\tilde{x}',\tilde{v}}{g,\tilde{n}=b\vdash x,\rho[\tilde{n_{i}}\mapsto\tilde{v}_{i}],x',v} & (Sbind)\\
\\
g,empty\vdash[],\rho,[],null & (Sempty) &  & g,any\vdash x:\tilde{x},\rho,\tilde{x},x & (Sany)\\
\\
\dfrac{isText(x)}{g,text\vdash x:\tilde{x},\rho,xs,x} & (Stext) &  & \dfrac{\begin{array}{c}
g(n)=n(\tilde{v})\vartriangleright b\\
g,b\vdash\tilde{x},[\tilde{v}\mapsto\tilde{e}(\rho)]\oplus\rho',\tilde{x}',v
\end{array}}{g,n(\tilde{e})\vdash x,\rho,x',v} & (Scall)\\
\\
\dfrac{\tilde{e}(\rho)=\tilde{v}}{g,\left[\tilde{e}\right]\vdash\tilde{x},\rho,\tilde{x},\tilde{v}} & (Ssynth) &  & g,ok\vdash\tilde{x},\rho,\tilde{x},null & (Sok)\\
\\
\dfrac{\begin{array}{c}
g,\gamma(g)\vdash\tilde{x},\rho\oplus(\rho'\backslash N),\tilde{x}',v\\
g(\rho\oplus(\rho'\backslash N))
\end{array}}{g,(t,N,\gamma,b)\vdash(t,\rho',\tilde{x}):\tilde{y},\rho,\tilde{y},v} & (Sel_{1}) &  & \dfrac{\begin{array}{c}
g,b\vdash\tilde{x},\rho\oplus(\rho'\backslash N),\tilde{x}',v\\
\neg\exists g\in dom(\gamma)\bullet g(\rho\oplus(\rho'\backslash N))
\end{array}}{g,(t,N,\gamma,b)\vdash(t,\rho',\tilde{x}):\tilde{y},\rho,\tilde{y},v} & (Sel_{2})
\end{array}
\]
\hfill{}

\protect\caption{Specification}
\label{fig:Specification}

\end{figure*}

The XML grammar language is used to specify XML languages. A grammar
defines a collection of XML trees; each tree is a member of the XML
language defined by the grammar. The association between an XML grammar
and a set of XML trees is defined as a relation of the form:

\[
g,b\vdash\tilde{x}+\tilde{x}',\rho,\tilde{x}',v
\]

\noindent where \emph{g} is the grammar, \emph{b} is a clause body,
$\tilde{x}$ is a sequence of XML elements, $\rho$ is an environment
associating variables with values, and v is a value. The relation
states that an XML document $d=doc(t,\rho,\tilde{x})$ satisfies a
grammar $g$ with starting rule named $n$ synthesizing value $v$
when $g,n()\vdash[(t,\rho,\tilde{x})],[],[],v$, i.e. calling the
rule named n with no arguments and in an empty variable environment
with respect to the root XML element must consume the complete element
and produce a value.

The relationship is defined in figure \ref{fig:Specification}. Rules
$SOr_{1}$ and $SOr_{2}$ specify the conditions under which a disjunction
recognises a sequence of XML trees. Two rules are required in order
to allow the recognition to succeed if either of the two patterns
succeed. The rule $Sand$ specifies the relationship between two patterns
\emph{in sequence}. The first pattern consumes a prefix of the sequence
of XML trees and passes the remaining trees to the second pattern.
The two binding environments associated with the individual patterns
are combined with $\oplus$ so that multiple occurrences of the same
variable name shadow on the right. This rule forces the binding for
(x=A)(y=B) to contain a binding for both x and y. It also forces the
environment for (x=A)(x=B) to contain a single binding for x that
is derived from B. The rule $Sbind$ describes the case in which variables
are bound to the result of recognizing a pattern. 

The rule $Sempty$ forces the sequence of XML trees to be empty and
synthesizes the null value. This is to be contrasted with the rule
$Sany$ that consumes a single tree. Empty can be used to force an
XML leaf element: $(X,[],[],empty)$ is a pattern that matches an
XML element with a tag X and with no children. This is to be contrasted
with $(X,[],[],any)$ that matches an XML element with tag X and a
single child element. The pattern $(X,[],[],any^{*})$ matches a single
tree with a tag X and with any number of children.

The rule $Stext$ recognizes a single XML text element. The rule $Scall$
is used to call a rule. Each rule may have more than one definition
in the grammar and has 0 or more arguments. The argument values are
supplied at the point of call and are expressions that are evaluated
with respect to the current variable bindings. The associations between
the formal parameters and the actual parameters form the initial environment
for the call. The result of the call is defined by the value produced
by the body of the clause.

The rule $Ssynth$ defines how values are synthesized. An action is
a known function. It is supplied with values that are constructed
by evaluating expressions in the context of an environment. The rule
describes the case where there is a sequence of expressions. This
allows a single pattern to return multiple values as in the following
rules:

\[
\begin{array}{c}
X()\vartriangleright[v,w]=Y()\:[v+w]\\
Y()\vartriangleright[10,20]
\end{array}
\]

\noindent where the rule X binds a pair of values v and w by calling
Y (which returns a pair of values 10 and 20). X terminates by returning
the sum of v and w (a single value).

The rule $Sel$ describes how XML elements are processed. An element
pattern involves a tag t, some attribute names A, some clauses consisting
of a guard and a pattern, and an otherwise pattern. Each guard is
a predicate that may reference variables whose values are bound in
the environment $\rho$. If the next XML element matches the required
tag and the children match a clause-pattern whose guard is satisfied
then the XML element is consumed and the value synthesized by the
clause-pattern is returned.

\section{Implementation\label{sec:Implementation}}

The previous section has specified how XML grammars can be used to
recognize an XML document and to synthesize a value in the process.
However the specification does not explain \emph{how} the parsing
process works. The aim of this paper is to explain how a SAX parser
can be made to efficiently parse an XML document with respect to a
grammar. 

Efficient parsing will be performed by translating the grammar into
a lookup table that predicts what to do based on the next SAX event.
Providing that the grammar has a specific property that makes each
lookup deterministic (the LL(1) property) then the table and SAX events
can be used to drive an efficient parsing machine.

To create the table from an XML grammar, the grammar must be translated
into a normal form. Section \ref{sub:Normal-Form} describes this
translation and section \ref{sub:Lookahead-Tables} defines an algorithm
that constructs the tables. Finally section \ref{sub:Parser} defines
a parsing machine.

\subsection{Normal Form\label{sub:Normal-Form}}

In order to process the grammar using a parsing engine it is necessary
to lift out all the disjunctions to the top level so that they become
alternative definitions for clauses. The following equivalence is
used to perform the transformation:

\[
G\cup\left\{ c(\tilde{m})\vartriangleright A(X|Y)B\right\} \equiv
\]

\[
G\cup\left\{ \begin{array}{c}
c(\tilde{m})\vartriangleright A(\tilde{n}=d(\tilde{v}))B\\
d(\tilde{v})\vartriangleright X\left\{ \tilde{n}\right\} \\
d(\tilde{v})\vartriangleright Y\left\{ \tilde{n}\right\} 
\end{array}\right\} 
\]

\noindent where $\tilde{v}=free(X|Y)$ and $\tilde{n}=bound(X|Y)$.
The idea is that any disjunction $X|Y$ makes reference to some variables
$\tilde{v}$ and binds some variables $\tilde{n}$. The disjunction
can be translated to a new clause with two alternative definitions
so long as the referenced variables are passed as arguments and the
bound values are returned as results. 

A simlar equivalence holds for element patterns:

\[
G\cup\left\{ c(\tilde{m})\vartriangleright A\:(t,N,{\displaystyle \bigcup_{i=1,n}}\tilde{g_{i}}\mapsto\tilde{b_{i}},b)\: B\right\} \equiv
\]

\[
G\cup\left\{ \begin{array}{c}
c(\tilde{m})\vartriangleright A(\tilde{n}=(t,N,{\displaystyle \bigcup_{i=1,n}}\tilde{g_{i}}\mapsto n_{i}(\tilde{v}_{i}),n(\tilde{w}))B\\
{\displaystyle \bigcup_{i=1,n}n_{i}(\tilde{v}_{i})\vartriangleright b_{i}[\tilde{w}_{i}]}\\
n(\tilde{w})\vartriangleright b[\tilde{n}]
\end{array}\right\} 
\]

\noindent The guarded patterns and else-pattern are transformed to
calls of new non-terminals. The free and bound variables are handled
in the same way as disjunction.

Repetition can be removed using the following equivalence:

\[
G\cup\left\{ c(\tilde{m})\vartriangleright AX^{*}B\right\} \equiv
\]

\[
G\cup\left\{ \begin{array}{c}
c(\tilde{m})\vartriangleright A(d(\tilde{v}))B\\
d(\tilde{v})\vartriangleright(x=X)(xs=d(\tilde{v}))[x:xs]\\
d(\tilde{v})\vartriangleright ok
\end{array}\right\} 
\]

\noindent The equivalences defined above are used left-to-right as
rewrite rules in order to transform XML grammars into a normal form
which is suitable for predictive parsing. The main aim is to get all
of the disjunctions lifted to the top-level of the grammar so that
calls can be indexed in terms of element tags. All the other transformations
support this aim by allowing variable bindings to be passed as arguments
in calls and the results of calls to be bound appropriately.

Consider the following grammar before transformation into normal form:

\begin{verbatim}
@Grammar Test
  A ::= <A> b = (B | C)* </A> {b}.
  B ::= <B n=name/> {n}.
  C ::= <C n=name/> {n}.
end 
\end{verbatim}

\noindent and after transformation:

\begin{verbatim}
@Grammar Test
  A ::= b = <A> C1 </A> {b}.
  C1 ::= x = C2 xs = C1 { Cons(x,xs) }.
  C1 ::= { Nil }.
  C2 ::= B.
  C2 ::= C.
  B ::= <B n = name> OK </B> {n}.
  C ::= <C n = name> OK </C> {n}.
end
\end{verbatim}

\subsection{Lookahead Tables\label{sub:Lookahead-Tables}}

Parsing is performed with respect to lookahead tables. Each clause
defines a lookahead table that maps element tags to sequences of patterns.
The lookahead table is constructed using the following clause properties:
\begin{lyxlist}{00.00.0000}
\item [{\emph{null}}] A clause is null if it is satisfied without processing
any XML elements.
\item [{\emph{first}}] The set of first tags associated with a clause.
A clause will process a sequence of XML elements. The first set of
a clause contains all tags for the head element of all such sequences.
If the first sets of a clause with alternative definitions are disjoint
for each definion then they can be used to predict which definition
to use.
\item [{\emph{follow}}] The set of follow tags associated with a clause.
A clause may be satisfied by an empty sequence of XML elements. On
completing the clause, the parse will continue to process a sequence
of XML elements. The follow set of a clause contains all tags for
the head element of such sequences, i.e. the XML tags that predict
no consumption of elements by a clause.
\end{lyxlist}
Section \ref{sub:Definition-of-Nullable} defines the \emph{null}
operation, section \ref{sub:Calculation-of-First} specifies an algorithm
that calculates the first and follow sets of grammar rules and finally
section \ref{sub:Table-Construction} shows how tables are constructed
and gives an example.

\subsubsection{Definition of Null\label{sub:Definition-of-Nullable}}

A clause element is \emph{null} when it can be parsed without consuming
any XML input. Predictive table construction uses the null property
to construct first and follow sets that are used to populate the table
for each gramar rule. The \emph{null} operation is defined by case
analysis on the elements as follows:\\

$\begin{array}{lcc}
null(n(\tilde{e}),g) & = & null(b,g),\: n(\tilde{v})\vartriangleright b\in g\\
null(b|b',g) & = & null(b,g)\vee null(b',g)\\
null(bb',g) & = & null(b,g)\wedge null(b',g)\\
null(\tilde{n}=b,g) & = & null(b,g)\\
null(empty,g) & = & true\\
null(any,g) & = & false\\
null(ok,g) & = & true\\
null(text,g) & = & false\\
null([\tilde{e}],g) & = & true\\
null((t,N,\gamma,b),g) & = & false
\end{array}$

\subsubsection{Calculation of First and Follow Sets\label{sub:Calculation-of-First}}

\begin{figure*}
\[
\begin{array}{ll}
(1) & \mathbf{repeat}\\
(2) & \quad\mathbf{for}\,(c(\tilde{v})\vartriangleright B)\in G\\
(3) & \qquad\mathbf{if}\,\forall b\in B\bullet null(b)\\
(4) & \qquad\mathbf{then}\, null(c)=true\\
(5) & \qquad\mathbf{end}\\
(6) & \qquad\mathbf{let}\, B'+\left\{ b\right\} +B''=B\\
(7) & \qquad\mathbf{in\,\mathbf{case}\,}b\mathbf{\,\mathbf{of}}\\
(8) & \qquad\qquad(t,\tilde{g}\mapsto\tilde{n},n)\,\mathbf{do}\\
(9) & \qquad\qquad\;\mathbf{for}\, x\,\mathbf{in}\, n:\tilde{n}\\
(10) & \qquad\qquad\qquad follow(x)=follow(x)\cup\left\{ \slash t\right\} \\
(11) & \qquad\qquad\;\mathbf{end}\\
(12) & \qquad\:\:\mathbf{end}\\
(13) & \qquad\:\:\mathbf{if}\,\forall b\in B'\bullet null(b)\\
(14) & \qquad\:\:\mathbf{then}\, first(c)=first(c)\cup first(b)\\
(15) & \qquad\:\:\mathbf{end}\\
(16) & \qquad\:\:\mathbf{if}\, isCall(b)\wedge\forall b\in B''\bullet null(b)\\
(17) & \qquad\:\:\mathbf{then}\, follow(b)=follow(b)\cup follow(c)\\
(18) & \qquad\:\:\mathbf{end}\\
(19) & \qquad\mathbf{\:\: let}\, D+\left\{ b'\right\} +E=B''\\
(20) & \qquad\mathbf{\:\: in\, if\,}isCall(b)\mathbf{\wedge\forall}b\mathbf{\in}D\bullet null(b)\\
(21) & \qquad\:\:\:\;\,\:\mathbf{then}\, follow(b)=follow(b)\cup first(b')\\
(22) & \qquad\:\:\:\;\,\:\mathbf{end}\\
(23) & \qquad\:\:\mathbf{end}\\
(24) & \qquad\mathbf{end}\\
(25) & \quad\mathbf{end}\\
(26) & \mathbf{until\,}not\, changed
\end{array}
\]
\protect\caption{Calculation of First and Follow Sets}

\label{fig:FirstAndFollow}
\end{figure*}

Calculation of the first and follow sets for the grammar is performed
by the algorithm defined in figure \ref{fig:FirstAndFollow}. The
rest of this section describes the algorithm.

The sets are calculated in a loop (1-26) that continues until a fixed
point is reached. Each clause in the grammar is processed in turn
(2). If every pattern in a clause named c is null then the clause
c is marked as null (4). For each pattern b in the body of the clause
(6), if the pattern is an element (8) then normal form has ensured
that the element clauses and the else pattern are all calls. Therefore,
all of the clauses called in the body of the element (9) are followed
by the tag t (10). If the prefix B' of the clause body is null (13)
then the clause c is predicted by the first set of b (14). If the
element b is a call and is followed by null patterns (16) then the
tags following b are the same as the tags following c. For all patterns
b' that occur after b in the clause body (19) if b is a call and the
intermediate patterns are null (20) then the tags following b are
those that predict b'.

A grammar is deterministic (or LL(1)) if there is at most one choice
at any given time. This is an important property because it makes
parsing efficient and relatively simple. Given a situation in which
a rule is called, if the grammar is deterministic then the next element
tag (as supplied by the SAX event mechanism) determines the grammar
rule to be used. If the grammar is not deterministic then more SAX
events have to be consumed in order to decide how to proceed or the
parsing machinery must support backtracking.

\subsubsection{Table Construction\label{sub:Table-Construction}}

\begin{figure*}
\[
\begin{array}{ll}
(1) & \mathbf{for}\, c(\tilde{n})\vartriangleright B+\left\{ b\right\} +B'\in G\qquad\mathbf{where}\:(\forall b\in B\bullet null(b))\wedge(first(b)\not=\emptyset)\\
(2) & \quad\mathbf{for}\, t\in first(b)\\
(3) & \qquad predict(c,t)=c(\tilde{n})\vartriangleright B+\left\{ b\right\} +B'
\end{array}
\]

\protect\caption{Definition of Predict}
\label{fig:Predict}

\end{figure*}

\begin{figure*}
\begin{tabular}{|c|c|c|c|c|c|c|}
\hline 
 & {\footnotesize{}A} & {\footnotesize{}/A} & {\footnotesize{}B} & {\footnotesize{}/B} & {\footnotesize{}C} & {\footnotesize{}/C}\tabularnewline
\hline 
\hline 
{\footnotesize{}A} & {\footnotesize{}b = <A> C1 </A>} &  &  &  &  & \tabularnewline
\hline 
{\footnotesize{}C1} &  & {\footnotesize{}\{ nil \}} & {\footnotesize{}x = C2 xs = C1 \{ cons(x,xs) \}} &  & {\footnotesize{}x = C2 xs = C1 \{ cons(x,xs) \}} & \tabularnewline
\hline 
{\footnotesize{}C2} &  &  & {\footnotesize{}B} &  & {\footnotesize{}C} & \tabularnewline
\hline 
{\footnotesize{}B} &  &  & {\footnotesize{}<B n = name> ok </B> \{n\}} &  &  & \tabularnewline
\hline 
{\footnotesize{}C} &  &  &  &  & {\footnotesize{}<C n=name> ok </C> \{n\}} & \tabularnewline
\hline 
\end{tabular}

\protect\caption{Predictive Parsing Table}
\label{fig:ExampleTable}
\end{figure*}

XML grammars are used to process XML documents using a predictive
parser. The parser processes a lookup table with respect to the grammar
and the next XML element. Each time a clause c is called in the grammar
with respect to an XML element with tag t, the relation predict(c,t)
is used to lookup the appropriate clause definition. The prediction
relation is defined in figure \ref{fig:Predict}.

Fortunately, it is easy to check whether an XML grammar is deterministic.
If the parse table contains at most a single entry in each cell, then
the grammar is LL(1). Only LL(1) grammars are supported by the parsing
machine defined in the next section.

Figure \ref{fig:ExampleTable} shows the lookup table corresponding
to the example defined in section \ref{sub:Normal-Form}. This table
has been produced by calculating the first and follow sets as defined
in figure \ref{fig:FirstAndFollow} and then populating the table
using the algorithm in figure \ref{fig:Predict}. Since all cells
have at most one entry, the grammar is LL(1), for example:

\[
predict(B,B)=B()\vartriangleright(B,\left\{ (n,name)\right\} ,ok)[n]
\]

\subsection{Parser\label{sub:Parser}}

\begin{figure*}
$\begin{array}{llcl}
(1) & (n(\tilde{e}):\tilde{p},\rho,\tilde{v},(t,\rho'):\tilde{x},d) & \longmapsto & ([b],\tilde{v}\mapsto\rho(\tilde{e}),\tilde{v},(t,\rho'):\tilde{x},(\tilde{p},\rho,d))\\
 &  &  & \qquad\mathbf{when}\: predict(n,t)=n(\tilde{v})\vartriangleright b\\
(2) & (n(\tilde{e}):\tilde{p},\rho,\tilde{v},\slash t:\tilde{x},d) & \longmapsto & ([b],\tilde{v}\mapsto\rho(\tilde{e}),\tilde{v},/t:\tilde{x},(\tilde{p},\rho,d))\\
 &  &  & \qquad\mathbf{when}\: predict(n,\slash t)=n(\tilde{v})\vartriangleright b\\
(3) & (n(\tilde{e}):\tilde{p},\rho,\tilde{v},text(t):\tilde{x},d) & \longmapsto & ([b],\tilde{v}\mapsto\rho(\tilde{e}),\tilde{v},text(t):\tilde{x},(\tilde{p},\rho,d))\\
 &  &  & \qquad\mathbf{when}\: predict(n,text)=n(\tilde{v})\vartriangleright b\\
(4) & ([],\_,\tilde{v},\tilde{x},(\tilde{p},\rho,d)) & \longmapsto & (\tilde{p},\rho,\tilde{v},\tilde{x},d)\\
(5) & ((t,N,{\displaystyle \bigcup_{i=1,n}}g_{i}\mapsto b_{i},b):\tilde{p},\rho,\tilde{v},(t',\rho'):\tilde{x},d) & \longmapsto & \begin{cases}
([b_{i},\slash t]+\tilde{p},\rho\oplus\rho',\tilde{v},\tilde{x},d) & \mathbf{when}\: t=t'\wedge g_{i}(\rho)\\
([b,\slash t]+\tilde{p},\rho\oplus\rho',\tilde{v},\tilde{x},d) & \mathbf{when}\: t=t'
\end{cases}\\
(6) & (\slash t:\tilde{p},\rho,\tilde{v},\slash t':\tilde{x},d) & \longmapsto & (\tilde{p},\rho,\tilde{v},\tilde{x},d)\;\mathbf{when}\; t=t'\\
(7) & ([\tilde{e}]:\tilde{p},\rho,\tilde{v},\tilde{x},d) & \longmapsto & (\tilde{p},\rho,\tilde{e}(\rho):\tilde{v},\tilde{x},d)\\
(8) & (empty:\tilde{p},\rho,\tilde{v},\slash t:\tilde{x},d) & \longmapsto & (\tilde{p},\rho,\tilde{v},\slash t:\tilde{x},d)\quad\mathbf{when}\:\tilde{p}=\slash t:\tilde{p}'\\
(9) & (any:\tilde{p},\rho,\tilde{v},text(t):\tilde{x},d) & \longmapsto & (\tilde{p},\rho,\bot:\tilde{v},\tilde{x},d)\\
(10) & (any:\tilde{p},\rho,\tilde{v},(t,\rho'):\tilde{x},d) & \longmapsto & (any(t):\tilde{p},\rho,\tilde{v},\tilde{x},d)\\
(11) & (any(t):\tilde{p},\rho,\tilde{v},\slash t:\tilde{x},d) & \longmapsto & (\tilde{p},\rho,\bot:\tilde{v},\tilde{x},d)\\
(12) & (any(t):\tilde{p},\rho,\tilde{v},(t',\rho'):\tilde{x},d) & \longmapsto & (any(t'):any(t):\tilde{p},\rho,\tilde{v},\tilde{x},d)\\
(13) & (any(t):\tilde{p},\rho,\tilde{v},text(t'):\tilde{x},d) & \longmapsto & (any(t):\tilde{p},\rho,\tilde{v},\tilde{x},d)\\
(14) & ((\tilde{n}=b):\tilde{p},\rho,\tilde{v},\tilde{x},d) & \longmapsto & ([b,\tilde{n}=]:\tilde{p},\rho,\tilde{v},\tilde{x},d)\\
(15) & ((\tilde{n}=):\tilde{p},\rho,\tilde{w}:\tilde{v},\tilde{x},d) & \longmapsto & (\tilde{p},\rho[\tilde{n}\mapsto\tilde{w}],\tilde{w}:\tilde{v},\tilde{x},d)\\
(16) & (text:\tilde{p},\rho,\tilde{v},text(t):\tilde{x},d) & \longmapsto & (\tilde{p},\rho,t:\tilde{v},\tilde{x},d)
\end{array}$

\protect\caption{Parsing Engine}
\label{fig:ParsingEngine}

\end{figure*}

A parse is performed using an engine that processes SAX events in
the context of a lookup table. The engine is defined using a state
transition function. The states of the engine are defined as follows:\\

$\begin{array}{lcll}
\sigma\in\Sigma & = & P\times\Phi\times[V]\times[S]\times D & states\\
p\in P & = & [B+I] & programs\\
i\in I & = &  & instructions\\
 &  & any(N) & any\: end\\
 & | & [N]= & bind\\
 & | & /N & tag\: end\\
s\in S & = &  & SAX\; events\\
 &  & N\times\Phi & start\: tag\\
 & | & /N & end\: tag\\
 & | & text(N) & text\\
d\in D & = &  & dumps\\
 &  & P\times\Phi\times D & call\: frame\\
 & | & \top & empty
\end{array}$\\

\noindent A machine state $(\tilde{p},\rho,\tilde{v},\tilde{x},d)$
consists of a program $\tilde{p}$ that is a sequence of clause elements
and machine instructions, an environment $\rho$ that associates variables
that are currently in scope with values, a stack of values $\tilde{v}$,
a sequence of SAX events $\tilde{x}$, and a dump $d.$ The idea is
that the program drives the machine. At each transition the next program
element and the current SAX event determines that happens. The current
context is saved on the dump when a grammar wule is called and then
the context is restored when the rule returns. Values are pushed onto
the value stack and, if the process terminates successfully then the
synthesized value is found at the head of the stack.

The machine executes with respect to an LL(1) lookup table that is
represented as a function $predict:N\times N\rightarrow C$ mapping
clause names and XML element tags to grammar clauses. Given an initial
call $c(\tilde{v)}$of a grammar rule, the machine uses a state transition
function to transform a starting state into a terminal state as follows:
\[
([c(\tilde{v})],[],[],[x],\top)\longmapsto^{*}([],[],[v],[],\top)
\]

\noindent If a terminal state cannot be reached then the parse fails.
The transition function is defined in figure \ref{fig:ParsingEngine}.
The machine is driven by case analysis at the head of the program.
Rules (1-3) define how a call is performed. The next SAX event is
either a start tag, an end tag or text. In each case the lookup table
is used to determine which rule is being called (the table cannot
be ambiguous and may contain no entry in which case the parse fails).
If the table contains an entry for the SAX event then the current
context is saved on the dump and a new context is created for the
execution of the rule body. Rule (4) shows what happens when a rule
body is exhausted; the saved context is restored.

Rules (5) and (6) show how element specifications are performed. When
an element specification is encountered in the program, a corresponding
SAX event to start an element must be received. In this case, either
one of the guard expressions is true, in which case the corresponding
body element is performed, otherwise the else-clause is performed.
In either case, a tag end instruction is added to the program which
will test for the corresponding end tag SAX event (6).

Rule (7) shows how actions are performed. The empty rule (8) defines
that children of XML element can be specified as empty.

The rules governing \emph{any} are defined (9 - 13). If an \emph{any}
element is encountered when the next SAX event is text then the text
is just ignored. If an \emph{any} element is encountered when the
next SAX event is a start tag then the corresponding end tag must
be consumed, therefore an \emph{any} machine instruction is created
to ensure these match up (10). Rules (11-13) define how the \emph{any}
instruction is processed for each type of SAX event.

Rules (14) and (15) define how binding ttakes place. When a \emph{bind}
element is encountered (14) the body element is added to the program
along with a \emph{bind} instruction. The \emph{bind} instruction
extends the environment with values in (15).

Finally, text is processed in rule (16).

\section{Analysis\label{sec:Analysis}}

This paper has specified and implemented a DSL for parsing XML documents
using the SAX event-based interface. The SAX interface is attractive
because it is efficient compared to the DOM interface which constructs
a model of the XML document before processing can start. The challenge
in processing SAX events is how to shield the user from implementation
details. Our approach is to use a DSL that allows XML languages to
be expressed as a standard grammar. This paper has provided a specification
and implementation of this language. The language has been implemented
as part of the XMF language oriented programming (LOP) \cite{LOP}
system which is open-source and available from \cite{XMF}. Further
details of the language can be found in \cite{Superlanguages}.

In addition, XMF can be used to export the grammars to an Java implementation
of the engine described in this paper. This allows XMF to be used
as a compiler for XML grammars that produce stand-alone XML parsers.
In these cases, the synthesizing actions are allows to be Java statements
and can be used to make calls on other APIs. This approach has been
used in a commercial context to process UML models encoded as XMI.

Originally, XML based languages were expressed in DTD-format and latterly
in XML schemas. \cite{Grammars} show that these formats can be expressed
using standard technology from formal language theory (i.e. language
grammars). The paper also investigates the properties of these grammars.

Kiselyov \cite{Functional} reports a number of XML parser implementations
in using declarative technologies including CL-XML (Common Lisp) \cite{CL-XML},
XISO (Scheme), Tony (OCaml) and HaXml (Haskell). As noted in \cite{Functional}
these are all DOM parsers and therefore suffer from the basic efficiency
problems inherent in DOM.

The parser reported in \cite{Functional} is implemented using a functional
style with many elegant features. However, it is not a true DSL for
XML parsing since it exposes the underlying implementation mechanisms.
The XML grammar language reported in this paper is implemented using
XMF which allows DSLs to be embedded within other languages.

\end{document}